\documentclass{article}

\usepackage{hyperref}
\usepackage[hyphenbreaks]{breakurl}
\usepackage{graphicx}
\usepackage{balance}
\usepackage{xspace}
\usepackage{subcaption}
\usepackage{comment}
\usepackage{csquotes}
\usepackage{color}
\usepackage[shortcuts]{extdash}

\newcommand{\wide}{{\it Large-DS}\xspace}
\newcommand{\focusedconsent}{{\it CookieOK-DS}\xspace}
\newcommand{\focusedbrowser}{{\it Browser-DS}\xspace}
\newcommand{\focusedlocation}{{\it Country-DS}\xspace}

\newcommand{\docker}{{CookieCheck}\xspace}
\newcommand{\testpage}{{WebPageTest}\xspace}
\newcommand{\directive}{{ePrivacy Directive}\xspace}
\newcommand{\cookiebar}{{Cookie Bar}\xspace}

\author{   Martino Trevisan$^\dagger$
           Stefano Traverso$^\dagger$$^\star$
           Hassan Metwalley$^\dagger$$^\star$
           Marco Mellia$^\dagger$$^\star$
           \\
           $^\dagger$Politecnico di Torino
           $^\star$Ermes Cyber Security SRL
           \\
           \texttt{\{first.last\}@polito.it }
\\
\\
{\color{blue}{\small An updated version of this study has been accepted at PoPETS 2019 }}
\\
{\color{blue}{\small and it is available at:}}
\\
{\color{blue}{\small \url{https://petsymposium.org/2019/files/papers/issue2/popets-2019-0023.pdf}}}
}
\title{Uncovering the Flop of the EU Cookie Law}

\begin{document}
\date{}
\maketitle

\begin{abstract}
In 2002, the European Union (EU) introduced the \directive to regulate the usage of online tracking technologies.
Its aim is to make tracking mechanisms explicit while increasing privacy awareness in users. 
It mandates websites to ask for explicit consent \textit{before} using any kind of profiling methodology, e.g., cookies. Starting from 2013 the Directive is mandatory, and now most of European websites embed a ``Cookie Bar'' to explicitly ask user's consent. 

To the best of our knowledge, no study focused in checking whether a website respects the Directive. For this, we engineer \docker, a simple tool that makes this check automatic.
We use it to run a measurement campaign on more than $35,000$ websites. Results depict a dramatic picture: 65\,\% of websites {\it do not respect} the Directive and install tracking cookies {\it before} the user is even offered the accept button.

In few words, we testify the failure of the \directive.  Among motivations, we identify the absence of rules enabling systematic auditing procedures, the lack of tools to verify its implementation by the deputed agencies, and the technical difficulties of webmasters in implementing it.

\end{abstract}

\section{Introduction}

When browsing the web, users encounter the so-called ``trackers''. They build their business on the massive collection and brokerage of personal data, and severely threaten users' privacy~\cite{hoofnagle2012privacy,turow2009americans}. 
To regulate the usage of tracking technologies in the web, the EU Commission published in 2002 the \directive, popularly known as the ``Cookie Law''. It is one of the most strict regulations on the usage of online tracking mechanisms~\cite{very_former_law,former_law,actual_law}. 
Article 5 requires websites to ask \textit{``prior informed consent for storage or for access to information stored on a user's terminal equipment''}. In other words, a website must ask the visitor if she authorizes the usage of cookies and similar tracking mechanisms (e.g., web beacons, Flash cookies, etc.) \textit{before} delivering and installing them.
Each EU member state has adopted the Directive starting from 2013. Since then, its effect has become evident to end users because of the presence of a ``Cookie Bar'' on most of websites (even if outside EU).

Despite the Directive has been in force for more than three years, there is no study reporting how it is actually implemented and to what extent.
In fact, to the best of our knowledge, there exists no tool to  automatically check if a website respects the Directive.
In this paper, we present \docker~\cite{cookie_check}, a simple tool we engineer to automatically perform this check. Given a website, \docker visits it as a ``new user'', and analyses cookies installed on the client browser. If any of them is classified as {\it profiling}, \docker tags the website as not compliant with the Directive.

We use \docker for a large measurement campaign involving websites from 25 countries and 25 categories. Results are discouraging: The large majority (65\%) of European websites violates the \directive, with some categories (e.g., ``News and Media'') where violations top to 92\%. 
Funnily, Adult sites come second in respecting it, after Law and Government sites.

Since each EU state has adopted a custom version of the \directive, we further check if users from countries with strictest (e.g., Italy) or not up-to-date (e.g., Germany) regulations get different protection from online tracking. Results show marginal changes. In a similar way, changing browser, or considering mobile/desktop devices has little impact on results. In short, the \directive is regularly violated.

The Directive has been criticized as a case of regulatory failure: it impairs user browsing experience, and it is ineffective in increasing the awareness about online tracking~\cite{koops2014trouble,markou2016behavioural,smart1,smart2}.
Here, we show that the Directive is a failure from the enforcement perspective too.
As possible causes, we identify the fact that the Directive deputizes each member state to monitor its enforcement without sketching standardized auditing procedures, as well as the technical difficulties for webmasters at monitoring the activity of third parties embedded in websites (see \autoref{sec:discussion} for a thorough discussion).



We strongly believe our results can be useful for both researchers and policymakers in the debate on how to regulate and enforce privacy policies on the web.


\section{Background}
\label{sec:background}

In this section we first describe the cookie technology in the context of online tracking.
Then, we describe the regulatory frameworks available in different countries to govern the usage of cookies.

\subsection{Cookies as tracking technology}
\label{sec:cookies}
Web services use cookies to store user information directly in the browsers. 
Each time the browser sends a request to a server, it can attach a cookie installed during previous communications, so that the status of the transaction is preserved over time.

When loading a web page the user's browser generally contacts different servers to fetch all page objects. Some of these are {\it first-party}, i.e., belonging to the {\it same domain} of the website, and some are {\it third-party}, i.e., belonging to {\it different domains}, e.g., Content Delivery Networks (CDN), advertisement platforms, etc. 
Cookies are installed on a per-domain basis.
Therefore, \textit{first-party cookies} can be installed by the website (the first party), whereas \textit{third-party cookies} can be installed by third-party servers. 



Cookies expiration time makes them also different. \textit{Session cookies} are temporary, i.e., deleted when the user closes the browser or after a short period of time.
\textit{Persistent cookies} contain an explicit expiration date and stay stored in the browser for long periods of time.

{\it Third-party persistent cookies} are the most prominent means used by trackers to reconstruct users' browsing trajectories and compute per-user profiles which are employed to, e.g., deliver personalized advertisement~\cite{englehardt_cookies}. As these threaten users' privacy, policymakers undertook initiatives to regulate their usage.

In this paper we focus on cookies installed by trackers, that we refer as \textit{profiling cookies}. Our approach can be extended to more advanced tracking techniques such as beacons, flash cookies, canvas, etc.

\subsection{Regulatory frameworks for web privacy}
\label{sec:regulation}

There exists no comprehensive regulatory framework concerning web privacy, and each country eventually provides its own -- see~\cite{data_protection_world} for an exhaustive comparison.

Some countries, e.g., Brazil, China and US, lack of a regulation concerning users' web privacy.
Other countries, e.g., Australia, India and Russia, developed regulatory frameworks 
that are vague with no technical details.
The most comprehensive frameworks which aim at regulating personal data collection and usage in the web are those provided by Canada, Switzerland and the EU. In this paper we focus on the latter, being its framework the most prominent, and involving a large number of countries.    

The EU introduced the Data Protection Directive~\cite{very_former_law} in 1995. It provided a definition for ``personal data'', 
and delineated criteria of transparency, legitimate purpose and proportionality.
It did not address web tracking, since, at that time, the web was at its birth.
The Directive has been amended in 2002~\cite{former_law} and 2009~\cite{actual_law}.
In the last version, it explicitly disciplines the use of any tracking ``devices'' (e.g., cookies, supercookies, fingerprinting, etc.), and it is based on the ``explicit consent'' principle.
It states that the website must i) provide a clear description of the entities wishing to install tracking devices, ii) install them only \textbf{after} explicit consent is provided by the user, and  iii) describe how the gathered information will be used.

Opinion 04/2012~\cite{opinion_exemption} explicitly considers cookies, and clarifies which kinds are exempted from the requirement of explicit consent: session cookies, and cookies that are {essential} to provide the service are exempted (e.g., to handle a cart in an e-commerce website). Prior consent is required for all others. The document explicitly observes that {\it third-party persistent cookies are typically not essential and, thus, not exempted}. This is the case of profiling cookies.
Opinion 15/2011~\cite{opinion_prior} details the consent procedure, and mandates the websites to i) offer the user a short resume of the privacy policy, ii) a link to page containing all details, and iii) an interactive element to explicitly ask users' consent to install tracking devices. All this information is typically provided in a ``Accept Cookie Bar'' offered the user the first time she visits the website.



All EU member states must transpose, and likely adapt, EU Directives into their regulatory body.
As such, EU countries has transposed the \directive in different ways.
For instance, France and Italy fully adopted the \directive, so that
French and Italian websites must require user explicit consent for cookies, except for those exempted. In Italy, websites not respecting the \directive can be fined up to $120,000$~Euro. 
Differently, Germany has not yet transposed the \directive to its regulatory body, with the result that German websites can simply describe the privacy policy adopted, but they are not subject to any fine for not asking explicit consent.

The \directive does not sketch procedures to guide the enforcement of its principles, nor provides guidelines to perform proper audits. 
\section{Measurement Approach}
\label{sec:meassetup}

Our analysis builds on automatic browsing of web pages. For this task, we use two different tools: \docker~\cite{cookie_check}, a custom tool we engineer to provide experiment parallelization, and \testpage~\cite{webpagetest} which enables higher configurability. 
Both take as input a set of web pages, and use a regular browser to visit each. When the page is fully loaded (i.e., the \texttt{OnLoad} event is triggered) or a $90$s timeout elapses, they dump to file the HTTP Archive (HAR)~\cite{har_spec}, a JSON-formatted structure that summarizes navigation data, and reports statistics about services and exchanged objects.
For each HTTP transaction, the HAR logs the headers of the corresponding request/response. If not otherwise specified, we take care of erasing the browser cache before each visit. No user action is performed on the page. Hence, we emulate the behavior of a new user accessing the web page for the first time, and not providing any consent to the installation of tracking devices such as cookies. 

To understand which web pages violate the \directive, we analyze the installed cookies at the end of each visit.
We first pick all HTTP responses with \texttt{Set-Cookie} header.\footnote{The \texttt{Set-Cookie} header contains the cookie the server generates to install on the browser.}
Next, we match the domain of the service installing the cookie against a list of $1,232$ well-known web-tracker domains, that we obtain from Better Privacy Tool~\cite{better_privacy}. Thus,
we consider only profiling cookies, i.e., third-party persistent cookies installed by trackers. For this, we pick those having a lifetime greater than a 1 month (details about threshold setting are provided in \autoref{sec:results_tpp}).

At the end of this process, we tag web pages violating the \directive as those installing at least one profiling cookie, i.e., (i) related to a web-tracking service, and (ii) persistent.
Our approach is conservative as it builds on subset of well-known tracking services, and does not include, e.g., advertisements platforms, CDNs, etc.
Second, we focus only on cookies leaving out all other tracking mechanisms.

\section{Tools and measurement campaigns}
\label{sec:dataset}

We run two measurement campaigns.
The first aims at checking the presence of profiling cookies at scale. 
The second investigates how tracking cookies differ depending on device, browser settings, country, and when consent has been given.

To perform these campaigns, we profit from a Linux machine equipped with an Intel\textsuperscript{\textregistered} Xeon CPU with 12 cores and 32GB RAM and connected to the Internet through a 1Gb/s network.
For the analysis of data we use a medium-size Hadoop cluster running Apache Spark and Python, whose \texttt{Cookie} module allows us to parse cookies.

\subsection{Large scale measurement campaign}
\label{sec:broad}
 
We rely on SimilarWeb~\cite{similarweb}, a web-site ranking service analogue to Alexa to obtain per-country and per-category website ranks. We pick popular websites in 21 EU member states, plus 4 extra-EU countries for a more comprehensive comparison.\footnote{SimilarWeb provides per-country ranks for 21 EU member states out of 28.}
We consider sites from 25 different categories. The list of countries and categories can be deduced from Fig.\ref{fig:cookie_ratio}. For each country and category, we pick the 100 most popular websites. In total, we consider $25 \times 25 \time 100$ services, corresponding to $36,197$ unique websites to visit as the lists partially overlap.

Performing measurements on such a wide number of websites is challenging on several points.
In particular, measurements must scale and be fast enough to visit websites in a reasonably short amount of time, without sacrificing accuracy. Hence, we build a custom web measurement tool, namely \docker, based on Google Chrome; its code can be downloaded at~\cite{dockercode}, and a working demonstrator of the tool is available online~\cite{cookie_check}. \docker runs in a Docker container that provides a reliable and lightweight means to isolate multiple browser instances.
24 Docker containers run in parallel to speed up the visiting cycles. We repeat each visit 5 times. In total we performed $180,985$ visits over a period of 15 days, and collected HAR files, corresponding to $213$\,GB of raw data. Less than 5\% of visits failed for timeout intervention, likely due to temporarily offline websites or slow loading speed.
In the remainder of the paper we refer to this dataset with \wide.\footnote{Given its size, we share this dataset on demand.}

\subsection{Specific measurement campaigns}
\label{sec:detailed}

We launch additional measurement campaigns using a different web measurement tool, \testpage from Google~\cite{webpagetest}. This tool 
offers a more flexible configuration than \docker, but it does not enable parallel testing, thus considerably increasing the collection time.
Hence, we employ \testpage to run experiments on smaller sets of web pages. In particular, we focus on the the 100 most popular websites for France, Germany and Italy.
In total we count 241 unique websites.
We then visit each of them, changing the configuration of the tool and mimicking different scenarios.

First, we evaluate the impact of providing consent to the installation of cookies. To this end, we visit the homepage of each website in our list, and manually give the consent to the usage of cookies whenever the \cookiebar is present.
We save the resulting browsing profile, and visit the websites again using \testpage.
This lets us verify whether websites actually install profiling cookies only upon user consent is provided.
We refer to this dataset as \focusedconsent.

Second, we investigate whether the use of different browsers may affect the number of installed cookies. For instance, mobile devices typically download simpler pages with less objects.
We run tests using all browsers available within \testpage: Internet Explorer, Mozilla Firefox and Google Chrome.
Thanks to Google Chrome advanced features, we emulate mobile browsers by changing both \texttt{User\-/Agent} and screen size;
we consider an Android smartphone and tablet of the Nexus series, and an Apple iPhone6 and iPad2. We call this dataset \focusedbrowser.

Finally, we study if and how websites change behavior when accessed 
from different countries. To comply with different local regulations, a website could react differently and install different sets of cookies depending on the client location (e.g, checking its IP address or language settings).
To validate this hypothesis we use HTTP-proxies to change the client IP address and the location of the the client.
For this experiment, we follow the same approach employed in~\cite{Marciel2016}. We use 8 proxies located in 8 different European countries, changing the locale settings accordingly.
In this case we use the desktop version of Google Chrome. 
We refer to this dataset as \focusedlocation.

\section{Results}
\label{sec:results}

Here, we first analyze cookie lifetime to define threshold for classifying profiling cookies. Then we unveil their usage in \wide. Finally, we study the effect on profiling cookies when providing consent, and changing browser or country.

\subsection{Distinguishing profiling cookies}
\label{sec:results_tpp}

\begin{figure}[t]
 \begin{center}
   \includegraphics[width=0.7\columnwidth]{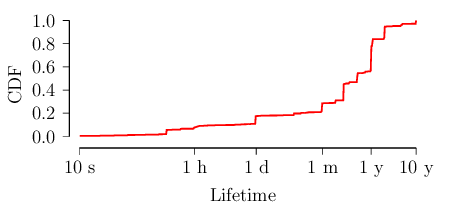}
   \caption{Empirical distribution calculated over lifetimes of third-party cookies in \wide.}
   \label{fig:expiration}
\end{center}
\end{figure}

\begin{figure*}[t]
  \begin{center}
    \includegraphics[width=\textwidth]{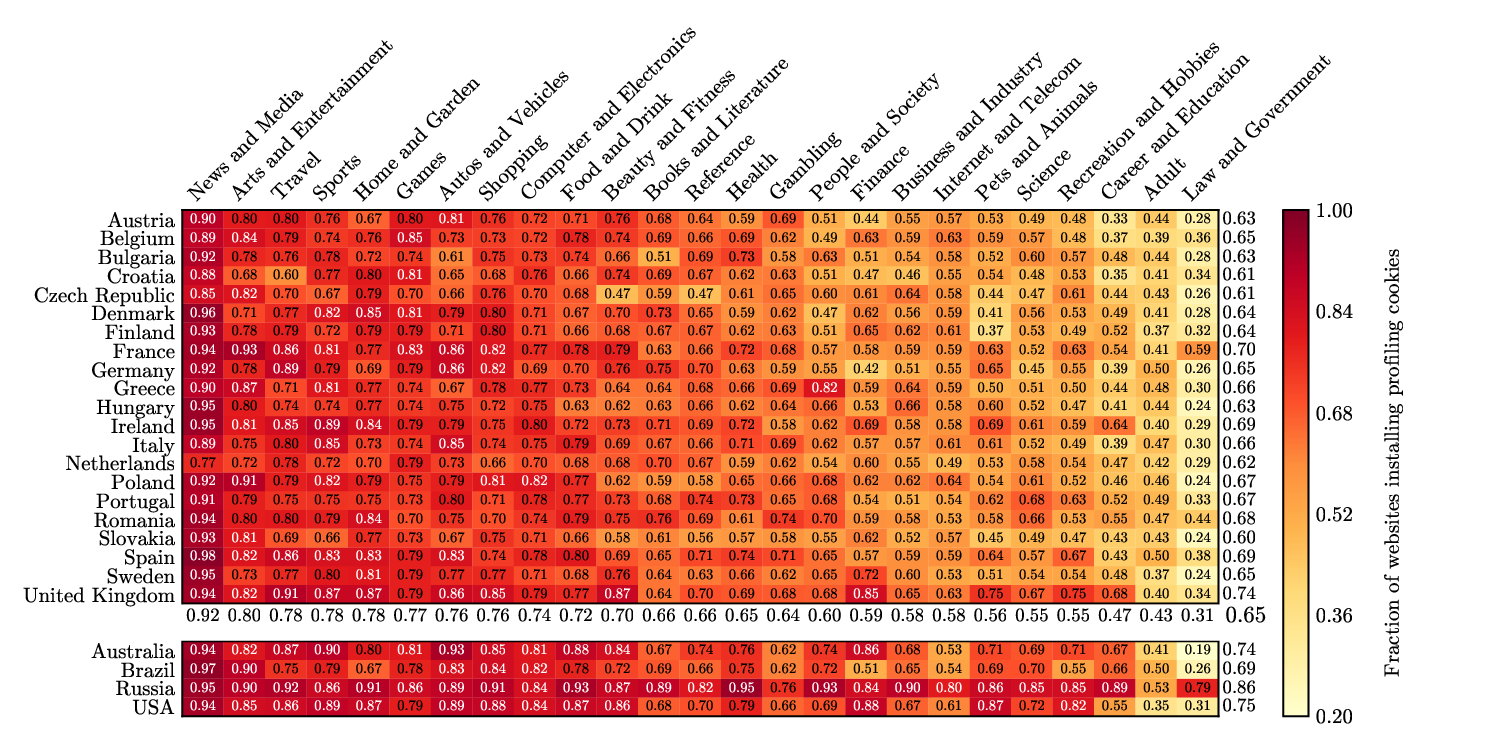}
    \caption{Per-country and per-category fractions of websites installing at least one profiling cookie in \wide. At the end of each row (column) it is reported the country-(category)wise average.} 
    \label{fig:cookie_ratio}
  \end{center}
\end{figure*}

As described in \autoref{sec:meassetup}, we consider as profiling a cookie installed by a tracking service and exhibiting a lifetime larger than a given threshold.
A similar approach was used by the authors of~\cite{englehardt_cookies}.
To define such threshold, we compute the cumulative distribution of third-party persistent cookie lifetimes in \wide. \autoref{fig:expiration} shows results. It clearly shows typical values, e.g., 1 day, 1 month, 1 year, and up to 10 years. 
As shown, 80\% of third-party cookies last 1 month or more. By manually inspecting them, we observe they belong mostly to web trackers and advertising services. 
Hence, to conservatively define profiling cookies, we consider all cookies installed by well-known tracking services (as by the Better Privacy Tool), whose expiration period is 1 month or higher. 

\subsection{Assessing \directive violations at scale}
\label{sec:assessing}

We leverage \docker to uncover the usage of profiling cookies across web pages for each country and category. For this, we use \autoref{fig:cookie_ratio}, where each cell reports the fraction of websites whereby the server installed at least one profiling cookie during any of the 5 visits. We remind that 100 websites are considered for each cell. Each row refers to a country, and each column to a category. 
At the end of each row we report the country-wise average fraction. Columns report the category-wise average fraction for EU countries only, and are sorted from the highest violator to the lowest.

Several observations hold. First, we notice that there exists no category whose fraction is close to 0. On average 66\% of websites violates the \directive. Recalling that our measurement is a conservative estimation (we neglect tracking mechanisms different from cookies), this fraction is startling.
Even the category with the lowest fraction of non-compliant websites -- ``Law and Government'' -- shows 31\% of websites that are not complaint with the Directive. This is possibly due to the fact that these websites usually do not embed a lot of advertisement since they do not build their business by selling online ads. Yet, third-party persistent cookies are installed -- possibly for analytics services. Funnily, Adult websites come second in being the more respectful of the Directive. They likely offer little (or specializes) ads, thus hosts fewer trackers.
On the opposite side, websites in ``News and Media'' are totally prone to violate the \directive (92\% of violations on average). This is no surprise as these websites typically hosts a lot of advertisers, and thus trackers.

If we consider per-country results, we observe that the picture is rather uniform, especially for EU member states. Values span from 60\% (Slovakia) to 74\% (UK). Interestingly, the former has not yet transposed the \directive to its regulation, while the latter did. Looking outside Europe, websites tend to install more profiling cookies at the first visit, with Russian (86\%) and US (75\%) websites being negative examples. This demonstrates the \directive is having some positive effect if compared to countries with no regulatory frameworks. However, this effect is small and far to be sufficient.

\begin{figure}[t]
 	  \begin{center}
 	    \includegraphics[width=0.7\columnwidth]{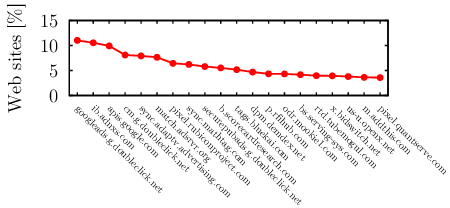}
 	    \caption{Pervasiveness of Top--20 verified trackers. Eleven are contained in more than 5\% of web sites.}
 	    \label{fig:tracker_pop}
 	  \end{center}
 	\end{figure}

To complement above results, we double check which trackers install cookies on users' browser without consent. 
\autoref{fig:tracker_pop} reports the pervasiveness of the most prominent trackers that install cookies before getting user's consent: Doubleclick from Google is the most present (appearing in 12\% of websites), followed by Adnxs by AppNexus (in 11\% of websites). The popularity of the trackers clearly influences the probability of finding it. 

In conclusion, there exist differences across website categories and countries, but a clear trend strongly emerges: 65\% of European websites ignore the \directive and let trackers install profiling cookies without any prior consent.

\subsection{Installed cookies upon consent}
\label{sec:results_banner}

\begin{figure}[t]
    \center

  \includegraphics[width=0.7\columnwidth]{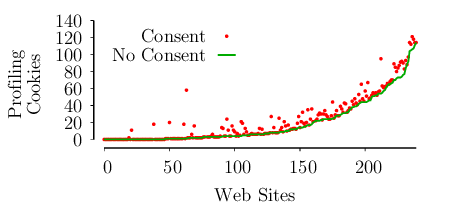}
   \caption{Per-website number of profiling cookies installed before and after giving consent.}
  \label{fig:first_second}
\end{figure}

\begin{table}[t]
  \centering
   \renewcommand{\arraystretch}{1.3} 
    \begin{tabular}{c|c|c}
      \textbf{Country} & \textbf{Banner} &  \textbf{Banner and Refresh}\\
        \cline{1-3} 
        France & 69 & 2 \\
        Germany & 31 & 0 \\
        Italy & 53 & 14 \\
    \end{tabular}
   \caption{Number of websites showing the \cookiebar,
            and refreshing the page upon user consent.}
   \label{tab:banner}
\end{table}

Websites that properly respect the Directive ask for consent by means of a button embedded in a \cookiebar. When clicked, the website refreshes (or updates) the page to deliver new and  enriched content with objects that trigger the installation of cookies.

We count the websites which implement this procedure, and thus do not violate the \directive. We consider three countries, as reported in \autoref{tab:banner}. Surprisingly, we observe that 54\% of websites in \focusedconsent do not even provide a \cookiebar, but regularly set tracking cookies!
In France and Italy, 69 and 53 out of 100 websites embed a \cookiebar, respectively.
For Germany, which has yet to transpose the Directive, only 31 websites do that.
Furthermore, the \cookiebar consent button triggers a page refresh in only 14 cases for Italian websites, just in 2 cases for French, and in no cases for German. 

We consider the same 241 website to count the number of profiling cookies which are installed \emph{before} and \emph{after} we provide consent. We report results in \autoref{fig:first_second},
where websites are sorted by the number of profiling cookies installed \emph{before} consent (green line). To this baseline we then sum new cookies installed \emph{after} consent has been given (red dots). As shown, only 43 websites do not install profiling cookies before obtaining consent.
All other websites, i.e., 80.5\% install profiling cookies before consent, and possibly install more after that.

\subsection{Impact of device and location}
\label{sec:results_browser_location}

\begin{figure}[t]
    \center
  \includegraphics[width=0.7\columnwidth]{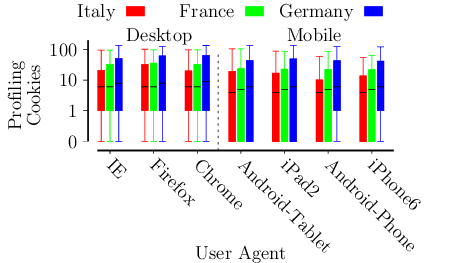}
  \caption{Number of profiling cookies set with different browsers.}
  \label{fig:user_agent}
\end{figure}

We study \focusedbrowser dataset to check whether installed profiling cookies vary when using different browsers, devices, or changing country where the client is located.

\autoref{fig:user_agent} reports the number of installed profiling cookies when using different browsers and devices, separately for Italian, French and German websites.
Box plots span from the $1^{st}$ to the $3^{rd}$ quartile, while whiskers report the $10^{th}$ and the $90^{th}$ percentiles; black strokes represent the median.
A quite large amount of profiling cookies is installed independently on the kind of browser or device, and the number only slightly decreases for mobiles. This is likely due to the simpler pages served to mobile devices.
Recall that Germany has not yet transposed the \directive at the moment of writing. This is reflected in the results with German websites installing more profiling cookies than French and Italian, respectively.

\begin{figure}[t]
   \begin{center}
     \includegraphics[width=0.7\columnwidth]{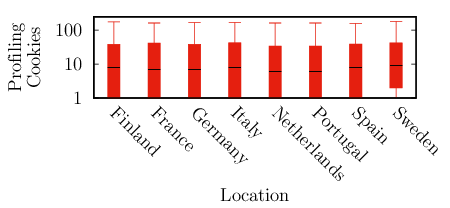}
    \caption{Number of profiling cookies set when changing crawling location. No significant difference is observed.}
    \label{fig:location}
   \end{center}
 \end{figure}

Finally, when the \focusedlocation websites are visited from 9 European countries (France, Italy, Germany, Finland, Netherlands, Portugal, Spain and Sweden)
the number of profiling cookies does not change. Results are reported in \autoref{fig:location} using boxplots, whose characteristics are the same as in the previous figure.
We conclude that websites do not adapt the set of profiling cookies to install if the country of the visitor implements the Directive in a different way.

\section{Related Work}
\label{sec:related}

To the best of our knowledge, this is the first technical work analyzing the adoption of the \directive at scale. 
Only Leenes and Kosta addressed the same problem~\cite{leenes2015taming}. They manually visited 100 Dutch websites and observed that most do not respect the \directive.
Similar observations emerged as a side effect in one of our previous studies~\cite{giannantoni_trackers_2017}.

Related to this work, it worths mentioning the work from Englehardt et al.~\cite{Englehardt:2016:OTM:2976749.2978313}. They addressed the problem of characterizing online tracking on very large number of websites. The dataset authors collect contains the same information we build upon to understand if a website violates the \directive, but they did not consider that the ``first visit'' may be different due to regulations in place. 

Some works studied the \directive from the legal perspective~\cite{koops2014trouble,markou2016behavioural}.
All concluded that this is a case of regulatory failure.
Koops~\cite{koops2014trouble} argues that the Directive creates the illusion that
individuals have control over their data, and it is difficult for webmasters to respect it.
Differently, Markou~\cite{markou2016behavioural} pinpoints that the cause of the failure lies in resistance of advertisement ecosystem and in the lack of privacy awareness among users.
Here, we quantify the failure by showing technical evidences rather than legal arguments.


\vspace*{-0.2cm}
\section{Discussion and conclusion}
\label{sec:discussion}


Despite being conservative, our results clearly uncover that the majority of websites ignores the \directive, testifying its flop.

We identify 5 main reasons behind this: \\
\noindent $i)$ the Directive does not offer guidelines to perform systematic auditing procedures with a coordinated supervision.
Two official reports from independent third parties~\cite{smart1,smart2} confirm this. In particular, \cite{smart2} emphasizes that enforcement of rules is ``\textit{insufficient and inconsistent}'', as currently in charge of each member state's competent authorities which ``\textit{tend to audit in cases where there is a specific risk or complaint by an individual. Ex-officio audits remain a minority.}''\\ 
\noindent $ii)$ The lack of automatic tools that can verify whether a website violates the Directive makes it possibly complicated for the deputed agencies to plan systematic audits.\\ 
\noindent $iii)$ The Directive does not delineate clear standardization procedures to uniform the composition of the \cookiebar and the implementation of the cookie consent acquisition. This lack complicates both being compliant with the Directive as well as its enforcement.\\
\noindent $iv)$ It is cumbersome for webmasters to fully support the Directive. Indeed, it is hard to control the activity of third parties, especially in the complicated ecosystem of advertisement and tracking platforms. Small websites cannot likely afford the additional technical and monetary costs.\\ 
\noindent $v)$ Despite users are becoming more and more conscious about their privacy being violated in the web, the overall consciousness level is still too low. Regulators should complement the \directive with proper awareness campaigns aiming at educating users about how their privacy is threaten under the surface of the web. Regarding this, the Regulatory Fitness and Performance (REFIT), the EU body in charge of verifying effectiveness of Directives, states the current rules end being counter-productive as ``\textit{the constant stream of cookie pop-up-boxes that users are faced with completely eclipses the general goal of privacy protection as the result is that users blindly accept cookies}''~\cite{refit_opinion}.

EU is currently drafting a revision of the Directive~\cite{revision_draft}. It aims to confirm the principles that govern the old one, but also to adapt them to new technologies and scenarios. Unfortunately, it still neglects above issues.



We have introduced \docker as a crude and simple tool to audit whether a website violates the \directive. It is available online~\cite{cookie_check}, along with source code~\cite{dockercode} and datasets collected for this study, so that it can be improved to include further checks.

Considering the research and measurement community, we are among the first to face the verification of privacy regulations. This is surprising, comparing to the effort given by the community to design tools to gauge, e.g., censorship and violations of network neutrality. We hope the effort in producing tools for auditing privacy violations will increase and get momentum.





\clearpage
\bibliographystyle{plain}
\bibliography{paper}

\end{document}